# A rigorous implementation of the Jeans–Landau–Teller approximation for adiabatic invariants

G. Benettin,* A. Carati,† G. Gallavotti‡

**Abstract:** *Rigorous bounds on the rate of energy exchanges between vibrational and translational degrees of freedom are established in simple classical models of diatomic molecules. The results are in agreement with an elementary approximation introduced by Landau and Teller. The method is perturbative theory "beyond all orders", with diagrammatic techniques (tree expansions) to organize and manipulate terms, and look for compensations, like in recent studies on KAM theorem homoclinic splitting.*

## 1. Introduction.

The problem of estimating energy exchanges between fast and slow degrees of freedom is a very relevant one in many domains of physics, and provides a main motivation for studying adiabatic invariants. Very elementary models come from gas dynamics, where the slow degrees of freedom correspond to the motion of the molecules center of mass, and the interaction with the fast internal degrees of freedom is provided by molecular collisions. In the simplest case one has just one vibrational and one translational degree of freedom; the Hamiltonian is:

$$H(x,y,p_x,p_y) = \frac{p_x^2}{2m} + V_0(x) + \frac{1}{2}(\frac{p_y^2}{\mu} + \mu\omega^2 y^2) + V_1(x,y) \; , \qquad (1.1)$$

and represents the collision of a diatomic molecule with a fixed wall, or with a point mass, in one dimension. Here $x$ and $p_x$ are the canonical coordinates of the center of mass, while $y$ and $p_y$ are the coordinates of the internal degree of freedom; $m$ is the total mass, $\mu$ the reduced mass and $\omega/2\pi$ is the vibration frequency. Both $V_0$ and $V_1$ are supposed to be smooth (in fact analytic) and to vanish for $|x| \to \infty$. One can take $V_1(x,0) = 0$; at fixed total energy one has then $|y| \sim \omega^{-1}$, and thus $V_1$ is small for large $\omega$, (generically) $V_1 \sim \omega^{-1}$.

Models like (1.1) have been introduced by Jeans [J1,J2] to investigate the times of relaxation to equilibrium in polyatomic gases, and to understand why in ordinary conditions, and in apparent conflict with the equipartition principle, the vibrational degrees of freedom of diatomic gases do not take part to the energy exchanges. In fact it is not well known that Jeans conjectured (after Boltzmann [Bo1,Bo2]) that the apparent "freezing" of fast vibrations, in this and other problems, is simply due to the fact that the energy exchanges with fast degrees of freedom are too small to produce any appreciable effect in a reasonable time scale (see [Be1,Be2,BGG1] for comments on this point). On the basis of heuristic arguments, Jeans proposed for the

---

* Università di Padova, Dipartimento di Matematica Pura e Applicata, Via G. Belzoni 7, 35131 Padova (Italy). E-mail: `benettin@pdmat1.math.unipd.it`.

† Università di Milano, Dipartimento di Matematica, Via Saldini 50, 20133 Milano (Italy). E-mail: `carati@mite35.mi.infn.it`.

‡ Università di Roma "La Sapienza", Dipartimento di Fisica "G. Marconi", Piazzale Aldo Moro 2, 00185 Roma (Italy). E-mail: `giovanni@ipparco.roma1.infn.it`.

energy exchanges an exponential law of the form

$$\Delta E \simeq \mathcal{E} e^{-\tau \omega} , \qquad (1.2)$$

where $\mathcal{E}$ and $\tau$ are natural units of energy and time (of the order of the energy per degree of freedom and of the collision time, respectively). As far as we know, this is the earliest proposal of an exponential law in connection with a problem of adiabatic invariance. Later Landau and Teller [LT] reconsidered the question in connection with the problem of sound dispersion, and worked out an elementary approximation scheme, actually not so different from Jeans' original one, which, once applied to models of molecular dynamics like the above (1.1), see [R], immediately leads to exponential laws like (1.2). We shall refer to this approach as to the Jeans-Landau-Teller (JLT) approximation. Besides the original papers [J1,J2,LT], the JLT approximation is described in [R,BCS,C].

To recall the essential result, following [BCS], we use for the fast degree of freedom the action–angle coordinates $I, \varphi$, and consider a general Hamiltonian like:

$$H(x, p, \varphi, I) = h(I) + \frac{p^2}{2m} + V_0(x, I; \varepsilon) + \varepsilon V_1(x, I, \varphi; \varepsilon) , \qquad p \equiv p_x . \qquad (1.3)$$

For a vibrational degree of freedom, which is the case in which we are mainly interested, one has $h(I) = \omega I$ and (generically) $\varepsilon \sim \omega^{-1}$. The coupling term $V_1$ can be conveniently taken to have zero average in $\varphi$, as the average could be added to $V_0$.

A possible formulation of the JLT approximation is as follows:

i) Ignore $V_1$, and solve the equations of motion for the reduced Hamiltonian $H_0(x, p; I) = \frac{p^2}{2m} + V_0(x, I)$, where $I$ is a parameter; let $\widehat{x}(t), \widehat{p}(t)$ be the solution for given initial data $(x^o, p^o)$ at $t = 0$.

ii) Use this solution, together with the free motion $\widehat{I}(t) = I^o$, $\widehat{\varphi}(t) = \varphi^o + \omega t$, where $\omega = \frac{\partial h}{\partial I}(I^o)$, in the Hamilton equation for $I$: $\dot{I} = -\varepsilon \frac{\partial V_1}{\partial \varphi}(\widehat{x}(t), \widehat{I}(t), \widehat{\varphi}(t); \varepsilon)$, integrate over time and compute $\Delta E = h(I(+\infty)) - h(I(-\infty))$.

As a result (see [BCS] for details), one gets $\Delta E$ in the form of a Fourier series in $\varphi^o$:

$$\Delta E = \sum_{\nu \in \mathbb{Z}} E_\nu \, e^{i\nu \varphi^o} \qquad (1.4)$$

with separate exponential laws for the different Fourier components: namely[1]

$$|E_\nu| = \mathcal{E}_\nu e^{-|\nu|\tau\omega} \quad \text{for } \nu \neq 0 , \qquad |E_0| = \mathcal{E}_0 e^{-2\tau\omega} \qquad (1.5)$$

(proportionality to a power of $\omega$ is possibly hidden in the constants $\mathcal{E}_\nu$); the crucial constant $\tau$ is given by the width of the analyticity strip of the Fourier coefficients of $V_1(\widehat{x}(t), \widehat{I}(t), \varphi; \varepsilon)$, thought of as functions of complex time $t$. In typical cases, and in particular for the potentials used in [LT,R], $\tau$ is the width of the analyticity strip of the approximate solution $\widehat{x}(t), \widehat{p}(t)$, and is easily computed. The precise identification of $\tau$, as well as the result that the average energy exchange $E_0$ decays much faster than the zero-average part (the latter practically coincides with the first Fourier term), are the essential achievements of the JLT approximation.

As shown in [BCS], the JLT approximation, although apparently crude, works indeed *excellently*, and in particular the separate exponential laws for the different Fourier components, as well as the identification of $\tau$, are fully confirmed by accurate numerical computations (the agreement between the theoretical value of $\tau$ and the numerically computed quantity is better than one percent, for relative energy exchanges in the rather wide range $10^{-4} \div 10^{-30}$). The reason of such a good agreement is not so clear; in particular (see [CBG] and [BCS] for comments) one does not understand why the analyticity strip $\tau$ of the zero–order solution $\widehat{x}(t), \widehat{p}(t)$ plays such a relevant role. One should note that, apart from exceptional cases, $\tau$ is *not* an approximation to the analyticity strip of the true solution: the latter is totally different from $\tau$, and even vanishes at large $\omega$ [BCS].

The purpose of this paper is to understand a simplified version of the problem (1.3) from a mathematically rigorous point of view. We shall restrict ourselves to the case of a harmonic oscillator, namely $h(I) = \omega I$ and to $I$–independent $V_1, V_0$, and work perturbatively with $\varepsilon$ as the perturbative parameter; only at the end we set $\varepsilon \sim \omega^{-1}$. Although our approach is in principle more general, we explicitly consider here the case $V_0(x)$ proportional to $e^{-x/l}$ where $l$ is a length scale; moreover, we shall assume that $V_1(x, \varphi, \varepsilon)$ factorizes as $g(\varphi)\tilde{V}_1(x)$, with $\tilde{V}_1(x)$ decaying sufficiently fast at $\infty$; we shall also occasionally refer, for definiteness, to $\tilde{V}_1$ proportional to $e^{-x/l}$. More general choices of $V_0$ and $V_1$ could be allowed at the only expense of a more involved notation: some comments on such generalizations will be made later, but we prefer to sacrifice the generality of the model to the exposition of a general technique in a rather special class of problems. We have not attempted, however, to replace the coupling strength $\varepsilon$ in (1.3) with a power of $\sqrt{I}$ and to take $V_0, V_1$ analytic in $\sqrt{I}$ (rather than $I$): the latter problem may require new ideas.

---

[1] Properly speaking, in the exactly isochronous case $h(I) = \omega I$, at first order in $\varepsilon$ the average $E_0$ vanishes: contributions proportional to $e^{-2\tau\omega}$ are however expected at higher orders.

Let us stress here that, although we treat explicitly the case of a single fast variable, our approach can be generalized to the case of several fast variables; for a preliminary result in this direction see Section 4.4.

Concerning the technique, we shall proceed as in [CG,G3,G1,CF1], and expand solutions in series of $\varepsilon$, by using a diagrammatic technique to represent $\Delta E$ and to look for relevant compensations. The first order of this expansion gives the JLT approximation, and due to the decay property of $V_0$ and $V_1$ at infinity, the perturbative series converge. In particular we obtain the Fourier coefficients $E_\nu$ of the energy exchange in the form of a series in $\varepsilon$: $E_\nu = \sum_{h \geq 1} \varepsilon^h E_{h,\nu}$, and (as is very crucial) all terms are bounded by the *same exponential factor*: precisely we get

$$E_{h,\nu} \sim \delta^{-hq+1} e^{-|\nu|\omega\tau(1-\delta)} \quad \text{for } \nu \neq 0 , \qquad E_{h,0} \sim \delta^{-hq+1} e^{-2\omega\tau(1-\delta)} , \qquad (1.6)$$

with some given $q > 0$ depending on the model, while $\delta > 0$ is a small free parameter. The leading terms $E_{1,\pm 1}$ are written in integral form. Denoting by $V_{1,\nu}$ the Fourier components of $V_1$ it is

$$E_{1,\pm 1} = \pm i \int_{-\infty}^{\infty} V_{1,\pm 1}(\widehat{x}(t), \widehat{I}(t)) e^{\pm i\omega t} \, dt \qquad (1.7)$$

In some interesting cases such terms can be explicitly evaluated: for example, for the very special choice (see [R])

$$V_0(x) \sim e^{-x/l} , \qquad V_1(x, \varphi, I) \sim e^{-x/l} \cos\varphi , \qquad l > 0 \qquad (1.8)$$

and in this case it is $E_{1,\pm 1} \sim \omega^2 e^{-\tau\omega}$. For such a model one finds $q = 4$ and $\tau = \pi/(2\sqrt{\eta^o})$, $\eta^o$ denoting the asymptotic value of the translational energy at time $-\infty$.

Inequalities (1.6) clearly solve Jeans' problem of giving an exponential upper bound on $\Delta E$. Indeed, taking for example $\varepsilon = 1/\tau\omega$ and making the choice $\delta = \varepsilon^\gamma$ with $0 < \gamma < q^{-1}$, by summing the series one gets that the first term dominates (for the values of $\varphi$ for which it is not zero) for small $\varepsilon$ and

$$|\Delta E| < \text{const.} \frac{e^{-\omega\tau_\omega}}{(\tau\omega)^{1-\gamma q}} , \qquad \tau_\omega = \tau[1 - (\tau\omega)^{-\gamma}] , \qquad (1.9)$$

so the equilibrium times are necessarily long (see [S] for preliminary results and comments on this point). Unfortunately, from the above analysis one cannot conclude that the first order in $\varepsilon$, namely the JLT approximation, dominates the expansion. Hence we cannot deduce from (1.6) a two–sided (or "exact") estimate for $\Delta E$: in fact the correction $1 - \delta$ at the exponent implies that the other (higher order) terms could exactly cancel the first order contribution, leading to a much smaller $\Delta E$; or they could even dominate leading to a $\Delta E$ much larger than the one of the JLT approximation (but of the order of $O(e^{-\tau\omega(1-\delta)})$, of course). A two–sided estimate can be obtained if one assumes $\varepsilon \ll \omega^{-q}$: in this case one can then take $\delta \sim \omega^{-1}$, and the correction $1 - \delta$ is then rigorously negligible.

*Remark:* In special cases, for instance with tidal *dipolar* forces in (1.1) (overall potential $V_0(x) + V_1(x, y) = V(x + \frac{1}{2}y) + V(x - \frac{1}{2}y)$, with some $V$), one has $V_1(x, y) = \mathcal{O}(y^2)$, and thus $\varepsilon \sim \omega^{-2}$, but this is not enough as one should have $\tilde{V}_1 = O(y^{2q})$ in order to be able to justify a choice of $\varepsilon$ of order $O(\omega^{-q})$ to be able find upper and lower bounds on $\Delta E$ with the above estimates (1.6). The only case in which (1.6) suffices for getting sharp upper and lower bounds seems to be the case in which the amplitude $\sqrt{I/\omega}$ of the oscillations is initially sufficiently small (one could see that $y$ remains then consistently small at any time). In such a situation it is reasonable to assume $\varepsilon$ to be small compared to $\omega^{-q}$, and deduce from (1.6) a lower estimate, too. This can be interpreted as indicating that an initially unexcited molecule may reach an energy of the order $\omega^{-2q}$ within a computable, exponentially large, number of collisions, (*i.e.* an exponentially large time). A proof of this fact is, however, not a well posed question in the frame of the above models because it requires studying models in which several collisions can be observed (while model (1.3) only considers one collision).

Our perturbative approach to the problem is quite different from the Nekhoroshev–like approaches to similar problems, see [N1,N2,BGG2,BF], which also lead to exponential estimates for the energy exchanges. Here we do not introduce chains of canonical transformations, rather (as in Lindstedt method) we expand solutions. The essential quantity, namely the exponential $e^{-\tau\omega}$, appears already at first order in $\varepsilon$, and not after optimizing a divergent series truncation. Moreover, the crucial constant $\tau$ is precisely identified, and associated with an easily computable parameter of the unperturbed Hamiltonian; the corresponding quantity, in the Nekhoroshev–like approach, appears instead indirectly, through the analysis of an apparently divergent rate of perturbative series, and a sharp determination of it seems practically impossible. We also remark that here the results are more detailed (one explains in particular the different exponentials in the bounds for the average and for the different Fourier components).

Our approach is instead rather close to the modern methods of investigating the problems of homoclinic splitting in forced pendula, and similar problems which are the basic steps to understand the Arnol'd diffusion, see [ACKR] and the review paper [G1]; some comments on this point are deferred to Section 4.3. A connection also exists with other versions of perturbation theory "beyond all orders", like the field theoretic approaches, see [PV,G2], and the modern KAM theory as started by Eliasson, see [E,G3,CF1]. Finally, our paper is quite close to [C], where one considers the JLT approximation scheme, and deduces an upper bound by a different technique, which essentially consists in a rigorous version of Jeans' original ideas.

For a study of a quantum problem related to the problem discussed here, see [JKP], [JP1], [JP2], [MN].

This paper is organized as follows: in Section 2 we introduce precise notations, and state formally our results, limiting ourselves to the restricted case described above. Section 3 is devoted to the proof, which essentially follows [G1,G3]; the only new ideas are the use of the energy–time coordinates, in place of $x,p$, for the slow degree of freedom, and the particular "cancellation mechanism" which at any order $h$ leads to the exponential bound (1.6) for $E_{0,h}$. *The latter is the only point* where the Hamiltonian character of the problem is used in an essential way. Section 4 shortly accounts for some possible generalizations of our approach: precisely, in Section 4.1 we extend the results to a wider class of perturbations; in Section 4.2 we show that, to get cancellations, the symplectic symmetry of the equations can be replaced by other symmetries, like the time reversal invariance of the equations; in Section 4.3 we establish the connection with the problem of separatrices splitting; in Section 4.4 we briefly treat the case of several fast variables; and in Section 4.5 we shortly indicate a possible way to improve our estimates in some special cases. Finally, in the Appendix we illustrate how one could proceed without the use of the energy–time variables.

## 2. Statement of results.

To stress the mathematical character of our analysis we assume that all physical quantities are expressed in natural dimensional units; in other words, from now on we deal with dimensionless quantities.

We consider the Hamiltonian:

$$H(x,p,\varphi,I) = \omega I + \frac{p^2}{2} + V_0(x) + \varepsilon g(\varphi)\widetilde{V}_1(x) \; ; \tag{2.1}$$

although we have in mind the general problem of scattering by a smooth potential barrier, we explicitly treat the case of $V_0$ as in (1.8): $V_0(x) = e^{-x}$. Concerning $\widetilde{V}_1$, we only need it be real analytic and to decay to zero for $x \to +\infty$; for definiteness we shall occasionally refer to the simple choice

$$\widetilde{V}_1(x) = e^{-x} \tag{2.2}$$

(such exponential potentials are taken from [LT,R]).

The one degree of freedom Hamiltonian

$$H_0(x,p) = \frac{p^2}{2m} + V_0(x) \; , \tag{2.3}$$

giving the motion of the center of mass for $\varepsilon = 0$, only admits unbounded (*i.e.* scattering) solutions, namely solutions with $x(t) \to \infty$ and $p(t) \to \pm p_\infty$ for $t \to \pm\infty$ for all energy values. We therefore introduce the

energy–time variables of the Hamiltonian (2.3), which are the analog of the common action–angle coordinates for unbounded motions. To this end, let us fix arbitrarily an origin $x^o$ for $x$, and consider motions with initial data $x^o$ and $p^o = \sqrt{2(\eta - V_0(x^o))}$, namely motions with total energy $\eta$. The corresponding solutions can be expressed as functions of $\eta$ and of time: which we shall denote by $\xi$. We write such motions as:

$$x = \widehat{x}(\xi, \eta), \qquad p = \widehat{p}(\xi, \eta). \tag{2.4}$$

For the above choice of $V_0$ one easily finds explicit expressions in terms of elementary functions, namely

$$\widehat{x}(\xi, \eta) = \log\left[\frac{1}{\eta}\left(\cosh\sqrt{\eta/2}(\xi - \xi^o)\right)^2\right], \qquad \widehat{p}(\xi, \eta) = \sqrt{2\eta}\tanh\sqrt{\eta/2}(\xi - \xi^o) \tag{2.5}$$

(with $\xi^o$ such that $\widehat{x}(0, \eta) = x^o$). In general, in any scattering problem, the solution (2.4) is implicitly defined just by a quadrature and an inversion. But the analysis of the analyticity properties may be involved and difficult: and this is the main reason for restricting our analysis to $V_0 = e^{-x}$.

Expressions (2.4) can be regarded as a change of variables from $(x, p)$ to $(\xi, \eta)$, which is, of course, a canonical change of coordinates. After the change of coordinates (2.4), the Hamiltonian $H_0$ becomes $K_0(\xi, \eta) = \eta$, while the Hamiltonian (2.1) is changed into:

$$K(\xi, \eta, \varphi, I) = \omega I + \eta + \varepsilon f(\xi, \eta) g(\varphi), \tag{2.6}$$

with

$$f(\xi, \eta) = \widetilde{V}_1(\widehat{x}(\xi, \eta)). \tag{2.7}$$

¿From such Hamiltonian one gets $\dot\varphi = \omega$, and thus $\varphi(t) = \varphi^o + \omega t$; for $\xi$, $\eta$ one then finds the pair of time–dependent equations

$$\dot\xi = 1 + \varepsilon f_\xi(\xi, \eta) g(\varphi^o + \omega t), \qquad \dot\eta = \varepsilon f_\eta(\xi, \eta) g(\varphi^o + \omega t), \tag{2.8}$$

with

$$f_\xi = \frac{\partial f}{\partial \eta}, \qquad f_\eta = -\frac{\partial f}{\partial \xi}. \tag{2.9}$$

The form (2.9) of $f_\xi$, $f_\eta$ expresses the Hamiltonian character of the problem. This however plays no role in the perturbative construction, and it is useful only at the very end, to see that many individually large additive contributions to the quantities that we want to estimate (*i.e.* energy exchanges between internal and external degrees of freedom) exactly compensate. We shall proceed with generic $f_\xi$ and $f_\eta$, and recall (2.9) only when necessary.

We now explain the assumptions we make on $g$, $f_\xi$ and $f_\eta$. Concerning $g$, we assume it to be analytic in a strip $|\text{Im}\,\varphi| < \rho$, with some $\rho > 0$, and bounded by

$$|g(\varphi)| \leq 1, \qquad \text{for} \quad |\text{Im}\,\varphi| < \rho \tag{2.10}$$

(taking one on the r.h.s. of the inequality is not restrictive). Properly speaking, in the special problem in (2.1) with perturbation independent of $I$, analyticity is not really necessary, and leads only to a minor improvement of the estimates. Analyticity is however necessary in the general case (see [N2] for a similar situation).

Concerning $f_\xi$ and $f_\eta$, let us preliminarily consider the Hamiltonian case, with $\widetilde{V}_1$ given by (2.2). In such a case the function $f$ defined by (2.7) is given by

$$f(\xi, \eta) = \frac{\eta}{[\cosh\sqrt{\eta/2}(\xi - \xi^o)]^2}, \tag{2.11}$$

and for any real $\eta$, $f$ is analytic, as function of $\xi$, in the complex strip $|\text{Im}\,\xi| < \tau(\eta) = \frac{\pi}{\sqrt{2\eta}}$. By moving $\eta$ in a complex ball around a real value $\eta^o$, the singularities of $f$ move, but $f$ remains analytic in domains like

$$\mathcal{D}_\delta(\eta^o) = \{(\xi, \eta) \in \mathbb{C} : |\eta - \eta^o| < c\delta\,;\, |\text{Im}\,\xi| < (1 - \delta)\tau\} \tag{2.12}$$

with $\tau = \tau(\eta^o)$, $\delta \leq 1$ and a suitably chosen $c$. Simple considerations lead to estimates on $f_\alpha$, $\alpha = \xi, \eta$:

$$\frac{1}{j!(m-j)!}\left|\frac{\partial^m f_\alpha}{\partial \xi^j \partial \eta^{m-j}}(s + i\sigma, \eta^o)\right| \leq C^m \delta^{-m-m_0} w(s) \qquad \text{for} \quad |\sigma| \leq (1-\delta)\tau(\eta^o) \tag{2.13}$$

where $m_0 = 3$, $C$ is a suitable constant, and $w$ is a suitable integrable function:

$$\int_{-\infty}^{\infty} w(s)\,\mathrm{d}s \;=\; W \;<\; \infty \qquad (2.14)$$

Under general assumptions on $f_\xi$, $f_\eta$, namely analyticity for $\eta$ in a small ball around a value $\eta^o$ and $\xi$ in a complex strip, and sufficiently fast decay (faster than $1/|\mathrm{Re}\,\xi|$) for $|\mathrm{Re}\,\xi| \to \pm\infty$, one gets estimates of the form (2.13), (2.14), with suitable $m_0$, $C$, $W$, and $\tau = \tau(\eta_0)$.

We shall use $m_0$, $C$, $W$ and $\tau$ as the quantities characterizing our problem, and express our results in terms of them. Let us note that, since $\xi$ is the time of the unperturbed motion, the quantity $\tau$ appearing in (2.13) is precisely the quantity appearing at exponent in the JLT approximation.

For $\varepsilon = 0$ the equations of motion (2.8) trivially admit the free solution $\xi_0(t) = t$, $\eta_0(t) = \eta^o$. For $\varepsilon \neq 0$ we are interested in real solutions such that, asymptotically for $t \to -\infty$,

$$\xi(t) - t \to 0\;, \qquad \eta(t) \to \eta^o \qquad (2.15)$$

Therefore we introduce the Taylor expansions (which will be proven to be convergent)

$$\xi(t) = t + \sum_{h \geq 1} \varepsilon^h \xi_h(t)\;, \qquad \eta(t) = \eta^o + \sum_{h \geq 1} \varepsilon^h \eta_h(t)\;, \qquad (2.16)$$

and rewrite this, in compact notation, as

$$\alpha(t) = \alpha_0(t) + \sum_{h \geq 1} \varepsilon^h \alpha_h(t)\;, \qquad \alpha = \xi \text{ or } \eta\;. \qquad (2.17)$$

The overall variation of $\alpha = \xi$ or $\eta$ (the asymptotic delay and respectively the energy exchange) is then $\Delta\alpha = \sum_{h \geq 1} \varepsilon^h \alpha_h(+\infty)$.

Denote by $\alpha_{h,\nu}$, $\nu \in \mathbb{Z}$, the Fourier components of $\alpha_h(+\infty)$, $\alpha = \xi$ or $\eta$, namely $\alpha_h(+\infty) = \sum_{\nu \in \mathbb{Z}} \alpha_{h,\nu}\, e^{i\nu\varphi^o}$; then the Fourier expansion of $\Delta\alpha$ is

$$\Delta\alpha = \sum_{\nu \in \mathbb{Z}} \Delta_\nu^\alpha\, e^{i\nu\varphi^o}\;, \qquad \text{with} \quad \Delta_\nu^\alpha = \sum_{h \geq 1} \varepsilon^h \alpha_{h,\nu}\;. \qquad (2.18)$$

Let us remark that the asymptotic energy exchange $\Delta E$ we discussed about in the Introduction, coincides with $-\Delta\eta$, and thus

$$E_{h,\nu} = -\eta_{\eta,\nu}\;, \qquad E_\nu = -\Delta_\nu^\eta = -\sum_{h \geq 1} \varepsilon^h \eta_{h,\nu}\;. \qquad (2.19)$$

We shall prove the following

**Proposition 1.** *Consider the equations of motion (2.8); let $\rho$, $m_0$, $C$, $W$ and $\tau$ be such that inequalities (2.10), (2.13) and (2.14) hold, for $0 < \delta \leq 1$, and consider motions satisfying the asymptotic conditions (2.15). Then:*

*i) For small $\varepsilon$ the Taylor series (2.17) for $\alpha(t)$ converge, precisely for any real $t$ one has*

$$|\alpha_h(t)| \leq AB^{h-1}\;, \qquad \alpha = \xi,\eta\;, \qquad (2.20)$$

*with suitable constants $A$, $B$ depending on $C$ and $W$.*

*ii) The Fourier components $\alpha_{h,\nu}$ of $\alpha_h(+\infty)$ satisfy the exponential estimates*

$$|\alpha_{h,\nu}| \leq AB^{h-1}\delta^{-hq+1}e^{-|\nu|\omega\tau(1-\delta)-\rho|\nu|}\;, \qquad \alpha = \xi,\eta \qquad (2.21)$$

*with $q = m_0 + 1$.*

*iii) In the Hamiltonian case (2.9), the average $\eta_{h,0}$ of the energy exchange satisfies the exponential estimate*

$$|\eta_{h,0}| \leq AB^{h-1}\delta^{-hq+1}e^{-2\omega\tau(1-\delta)} \qquad (2.22)$$

iv) *At first order in $\varepsilon$ one has the explicit expression*

$$\alpha_{1,\nu} = g_\nu \int_{-\infty}^{\infty} f_\alpha(t, \eta^o) e^{i\nu\omega t} dt , \qquad \nu \in \mathbb{Z} . \tag{2.23}$$

*Possible values of the constants $A$ and $B$ are $A = W$, $\qquad B = 8CW$.*

Point iv) of the proposition reports the expression of the Fourier coefficients of $\Delta\xi$, $\Delta\eta$ at first order in $\varepsilon$; as will be clear from the proof, one could produce explicit expressions for higher orders, too. For the choice (2.2) of $\widetilde{V}_1$, one has in particular

$$\eta_{1,1} = 4\pi i \, g_1 \, \frac{\omega^2}{e^{\omega\tau} - e^{-\omega\tau}}, \qquad \Delta E = 8\pi \frac{\omega^2}{e^{\omega\tau} - e^{-\omega\tau}} \sin\varphi^o + \ldots \tag{2.24}$$

with $\Delta E$ determined up to corrections of order $O((\varepsilon\omega^q)e^{-2\omega\tau})$.

### 3. Proof of Proposition 1.

*3.1. The recurrent scheme, and proof of item i).*

By introducing in the equations of motion (2.8) the Taylor expansions (2.16), collecting then terms of the same order $h$ in $\varepsilon$, one gets

$$\dot{\alpha}_h(t) = F_{\alpha,h}(t) , \qquad \alpha = \xi, \eta , \quad h \geq 1 , \tag{3.1}$$

with

$$F_{\alpha,1}(t) = g(\varphi^o + \omega t) f_\alpha(t, \eta^o) , \tag{3.2}$$

and for $h > 1$

$$F_{\alpha,h}(t) = g(\varphi^o + \omega t) \sum_{1 \leq m \leq h-1} \sum_{0 \leq j \leq m} f_\alpha^{m,j}(t, \eta^o) \sum_{\substack{k_1,\ldots,k_m \geq 1 \\ |k| = h-1}} \xi_{k_1}(t) \cdots \xi_{k_j}(t) \eta_{k_{j+1}}(t) \cdots \eta_{k_m}(t) \tag{3.3}$$

where use has been made of $\xi_0(t) = t$, and

$$f_\alpha^{m,j} = \frac{1}{j!(m-j)!} \frac{\partial^m f_\alpha}{\partial \xi^j \partial \eta^{m-j}} , \qquad |k| = \sum_i k_i .$$

Equation (3.1) is trivially solved by

$$\alpha_h(t) = \alpha_h(0) + \int_0^t F_{\alpha,h}(t') \, dt' .$$

The value of $\alpha_h(0)$ is determined by imposing the asymptotic condition (2.15); one gets

$$\alpha_h(t) = \int_{-\infty}^t F_{\alpha,h}(t') \, dt' . \tag{3.4}$$

Equations (3.2), (3.3) and (3.4) provide a recurrent scheme to compute $\xi(t)$, $\eta(t)$ to all orders in $\varepsilon$.

We can now deduce the estimates (2.20), which in turn imply the convergence of the series (2.16). Using the assumption (2.13) for $\delta = 1$, as well as $|g(\varphi)| \leq 1$, and denoting

$$\zeta_h = \sup_{t \in \mathbb{R}} \max \left( |\xi_h(t)|, |\eta_h(t)| \right) ,$$

we immediately get the recursive estimates

$$\zeta_1 \leq W, \qquad \zeta_h \leq W \sum_{0 \leq m \leq h-1} (m+1) C^m \sum_{\substack{k_1, \ldots, k_m \geq 1 \\ |k| = h-1}} \zeta_{k_1} \cdots \zeta_{k_m} ; \qquad (3.5)$$

the estimates (2.20) are then an immediate consequence of the following lemma:

**Lemma 1.** *The numbers $z_h$, $1 \leq h < \infty$, defined recursively by*

$$z_h = a \sum_{1 \leq m \leq h-1} b^m \sum_{\substack{k_1, \ldots, k_m \geq 1 \\ |k| = h-1}} z_{k_1} \cdots z_{k_m}, \qquad z_1 = c \qquad (3.6)$$

*satisfy the inequality*

$$z_h \leq \frac{c \, [4b \max(a, c)]^{h-1}}{h}. \qquad (3.7)$$

To prove (2.23) one applies this lemma with $c = a = W$ and $b = 2C$; $m+1$ in (3.5) is overestimated by $2^m$. One immediately gets $\zeta_h \leq z_h \leq W(8CW)^{h-1}/h$, even stronger than (2.20).

Lemma 1 is in turn an easy consequence of the following combinatorial Lemma 2, which also will be useful in the following.

**Lemma 2** *The numbers $M_h$, $1 \leq h < \infty$, defined recursively by*

$$M_h = \sum_{1 \leq m \leq h-1} \sum_{\substack{k_1, \ldots, k_m \geq 1 \\ |k| = h-1}} M_{k_1} \cdots M_{k_m}, \qquad M_1 = 1, \qquad (3.8)$$

*are given by*

$$M_h = \frac{2^{h-1}(2h-3)!!}{h!} < \frac{4^{h-1}}{h} \qquad (3.9)$$

Lemmas 1 and 2 are left to the reader with the hint that (3.8) implies for the generating function $u(x) = \sum_{h \geq 1} M_h x^h$ that it has to verify the relation $u(x)^2 = u(x) - x$, $u(0) = 0$, which is solved by $u(x) = \frac{1}{2}(1 - \sqrt{1-4x})$.

### 3.2. The tree expansion, and proof of points ii) and iv).

Equation (3.4), with $F_\alpha^h$ given by (3.3), gives $\alpha_h(t)$ as a sum of integrals, one for each choice of $m$, $j$ and $k_1, \ldots, k_m$, within the range specified in the corresponding sums. To each term one naturally associates an *elementary tree* (see figure 1), namely a *root* which ends in a *vertex* $v$ and there divides into $m$ *branches*; the root has a label $t \in \mathbb{R}$ and a label $\alpha = \xi$ or $\eta$, while the branches carry labels $\alpha_{k_1}, \ldots, \alpha_{k_m}$ (i.e., either $\xi_{k_i}$ or $\eta_{k_i}$); the rule is that "$\xi$-type" branches must precede (top to bottom, say) "$\eta$-type" branches. The labels $\alpha$ and $t$ attached to the root, together with $h = 1 + \sum_i k_i$, indicate that the graph is a contribution to $\alpha_h(t)$; the vertex represents an integration over a dummy variable $t_v$, in the range $-\infty < t_v \leq t$; the function to be integrated is the product of several factors, namely one factor $\xi_{k_i}(t_v)$ or $\eta_{k_i}(t_v)$ for each outcoming branch, and an additional factor $g(\varphi^o + \omega t_v) f_\alpha^{m,j}(t_v, \eta^o)$ for the vertex, $j$ being the number of $\xi$-type branches. For example, the elementary tree drawn in figure 1 represents a contribution to $\xi_8(t)$, namely

$$\int_{-\infty}^t g(\varphi^o + \omega t_v) f_\xi^{3,2}(t_v) \xi_2(t_v) \xi_1(t_v) \eta_4(t_v) \, dt_v . \qquad (3.10)$$

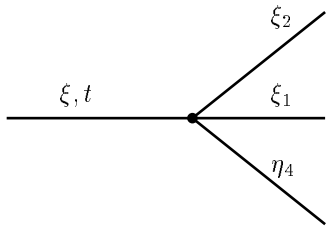

*Figure 1*. An elementary tree contributing to $\xi_8(t)$.

¿From elementary trees one then constructs *trees*, just by recursively expanding all branches in elementary trees; the procedure stops when the end branches represent either $\xi_1$ or $\eta_1$. The latter functions, in turn, are expressed via (3.4) and (3.2) as integrals of $g$ and $f_\xi$ or $f_\eta$. Each tree clearly represents one term of the sum obtained by recursively expanding each of the $\xi_{k_i}$ or $\eta_{k_i}$, appearing in (3.3), via (3.4) and (3.3) or (3.2), till everything is explicitly expressed in terms of $g$, $f_\xi$, $f_\eta$ and of the Taylor coefficients of $f_\xi$, $f_\eta$. See figure 2 for an example, actually a possible expansion of the elementary tree drawn in figure 1. Each tree $\theta$ with $h$ vertices (including the end vertices) has labels $t \in \mathbb{R}$ and $\alpha = \xi$ or $\eta$ on the root, indicating that it represents a contribution to $\alpha_h(t)$; each branch in the tree has in turn a label $\alpha = \xi$ or $\eta$, and the rule is that $\xi$–type branches must precede $\eta$–type branches issuing from the same vertex. The number of branches issuing from $v$ is denoted $m_v$, while the number of $\xi$–type branches is denoted $j_v$ ($j_v = m_v = 0$ for end vertices).

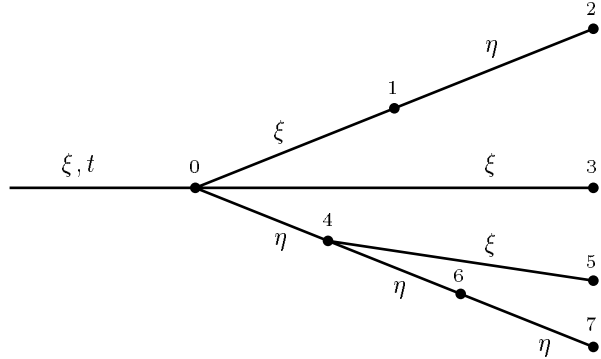

*Figure 2.* A possible expansion of the elementary tree drawn in Figure 1. Labels of branches and numbering of vertices, corresponding to the multiple integral below, are reported in the drawing.

The contribution, or *value*, of a tree is obtained as follows. Any vertex $v \in \theta$ represents an integration over a dummy variable $t_v$, so for each tree with $h$ vertices one has a multiple integral of multiplicity $h$. The quantity to be integrated is the product of one factor for each vertex $v \in \theta$, precisely

$$g(\varphi^o + \omega t_v) f_{\alpha_v}^{m_v, j_v}(t_v, \eta^o) \tag{3.11}$$

($g(\varphi^o + \omega t_v) f_{\alpha_v}(t_v, \eta^o)$ for end vertices), $\alpha_v$ being the label of the branch ending in the vertex $v$. Finally, the domain of integration respects the partial ordering[2] of vertices in the tree: if $\underline{t} = (t_0, \ldots, t_{h-1}) \in \mathbb{R}^h$ is the set of integration variables, then the domain is

$$T_t(\theta) = \{\underline{t} = (t_0, \ldots, t_{h-1}) \in \mathbb{R}^h : t_v \le t_{v'} \text{ if } v' \text{ precedes } v \, ; \, t_0 \le t\} \, .$$

For example, the contribution of the tree drawn in figure 2 is

$$\int_{-\infty}^{t} dt_0 \int_{-\infty}^{t_0} dt_1 \int_{-\infty}^{t_1} dt_2 \int_{-\infty}^{t_0} dt_3 \int_{-\infty}^{t_0} dt_4 \int_{-\infty}^{t_4} dt_5 \int_{-\infty}^{t_4} dt_6 \int_{-\infty}^{t_6} dt_7 \, g(\varphi^o + \omega t_0) \cdots$$
$$\cdots g(\varphi^o + \omega t_7) f_\xi^{3,2}(t_0) f_\xi^{1,0}(t_1) f_\eta(t_2) f_\xi(t_3) f_\eta^{2,1}(t_4) f_\xi(t_5) f_\eta^{1,1}(t_6) f_\xi(t_7)$$

(the way the integrals are nested reflects the tree structure). In fact, our aim is to compute $\alpha_h(+\infty)$, and to this purpose we must extend the outermost integration to $+\infty$; this means that the domain of integration is $T_\infty(\theta)$.

We shall denote by $\Theta_h$ the set of topologically distinguishable trees with $h$ vertices (the order in which the branches appear from top to bottom is relevant, so the trees of figure 3 are different); the set of labels, namely $\underline{\alpha} = \{\alpha_v, \, v \in \theta\}$ (the first component is the label $\alpha$ of the root) constitute the *decoration* of the tree. To prevent confusion, let us stress that $\underline{\alpha}$ refers to the whole set of labels on the branches, including the root, while $\alpha$ refers to the root, and indicates the quantity one wants to compute. For each tree, the decoration $\underline{\alpha}$ determines in particular $\underline{j} = \{j_v, \, v \in \theta\}$. Having introduced this formalism, we can write

$$\alpha_h(+\infty) = \sum_{\theta \in \Theta_h} {\sum_{\underline{\alpha}}}' V(\theta, \underline{\alpha}) \, ,$$
$$V(\theta, \underline{\alpha}) = \int \cdots \int_{T_\infty(\theta)} \prod_{v \in \theta} g(\varphi^o + \omega t_v) f_{\alpha_v}^{m_v, j_v}(t_v, \eta^o) \, d\underline{t} \, ; \tag{3.12}$$

---

[2] Any tree–like graph clearly induces a partial ordering in the set of vertices (from left to right in our figures).

the primed sum over $\underline{\alpha}$ is intended to run over all allowed choices of the branch decoration, of course not on the label $\alpha$ of the root. The quantity $V(\theta, \underline{\alpha})$ represents the value of the decorated tree $(\theta, \underline{\alpha})$. Let us remark that the set of trees for $\xi_h$ and for $\eta_h$ differ only by the label $\alpha = \xi$ or $\eta$ on the root.

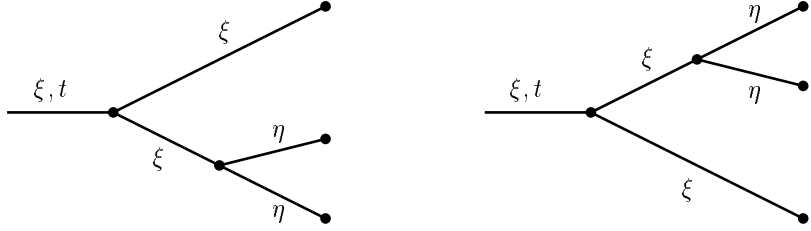

*Figure 3.* Different trees contributing to $\xi_5(t)$, to be counted separately.

To separate the different Fourier components $\alpha_{h,\nu}$ of $\alpha_h(+\infty)$, we must expand each function $g$ appearing in (3.12) in Fourier series, and conveniently collect terms. To this end, we add further decorations to the tree, namely we attach to each vertex $v \in \theta$ a label $n_v \in \mathbb{Z}$, corresponding to the chosen Fourier component of $g$, and collect decorated trees such that $\sum_v n_v = \nu$ (trees with *momentum m*, in the terminology of [G1], [G2]). So we write

$$V(\theta, \underline{\alpha}) = \sum_{\nu \in \mathbb{Z}} e^{i\nu \varphi^o} \sum_{\underline{n} \in \mathbb{Z}^h_\nu} V(\theta, \underline{\alpha}, \underline{n}) , \qquad \mathbb{Z}^h_\nu = \{\underline{n} \in \mathbb{Z}^h : \sum_v n_v = \nu\} ,$$

with

$$V(\theta, \underline{\alpha}, \underline{n}) = \left(\prod_{v \in \theta} g_{n_v}\right) \int \cdots \int_{T_\infty(\theta)} \left[\prod_{v \in \theta} f^{m_v, j_v}_{\alpha_v}(t_v, \eta^o)\right] e^{i\omega \underline{n} \cdot \underline{t}} \, d\underline{t} \qquad (3.13)$$

(the dot at the exponent denotes the usual Euclidean scalar product). Correspondingly one has

$$\alpha_{h,\nu} = \sum_{\theta \in \Theta_h} {\sum_{\underline{\alpha}}}' \sum_{\underline{n} \in \mathbb{Z}^h_\nu} V(\theta, \underline{\alpha}, \underline{n}) . \qquad (3.14)$$

We can now proceed to the estimate of $\alpha_{h,\nu}$, in order to prove point ii) of the proposition. To this end we consider the multiple integration on $\underline{t}$ in (3.13), and simultaneously raise all integration paths to $\text{Im}\, t_v = \pm \tau(1-\delta)$, with sign independent of $v$ and coinciding with the sign of $\omega \nu$. The only "vertical" paths one introduces in this way are at infinity, so they give no contribution to the integral; the analyticity of all functions in the strip $|\text{Im}\, t| \leq \tau$, implies that the value of $V(\theta, \underline{\alpha}, \underline{n})$ does not change. Having raised the integration paths, one immediately gets the exponential factors one is looking for, since one has

$$e^{i\omega \underline{n} \cdot \underline{t}} = e^{-|\nu|\omega \tau(1-\delta)} e^{i\omega \underline{n} \cdot \underline{s}} , \qquad \underline{s} = \text{Re}\, \underline{t} .$$

The estimate of $\alpha_{h,\nu}$ proceeds then as follows:

a) Using (2.10) one gets

$$\left| \sum_{\underline{n} \in \mathbb{Z}^h_\nu} \prod_{v \in \theta} g_{n_v} e^{i\omega \underline{n} \cdot \underline{s}} \right| \leq e^{-\rho|\nu|} .$$

Indeed, the r.h.s. is the $\nu$-th Fourier coefficient of $\mathcal{G}(\varphi) = \prod_{v \in \theta} g(\varphi + \omega s_v)$; such a function is clearly analytic in the strip $|\text{Im}\,\varphi| < \rho$, and is there bounded by $|\mathcal{G}(\varphi)| \leq 1$, so the estimate follows.[3]

b) From (2.13), by using the relation $\sum_v m_v = h - 1$, one gets

$$\prod_{v \in \theta} |f^{m_v, j_v}_{\alpha_v}(t_v, \eta^o)| \leq C^{h-1} \delta^{-hq+1} \prod_{v \in \theta} w(\text{Re}\, t_v) ;$$

---

[3] In the non analytic case, one would simply get one, in place of $e^{-\rho|\nu|}$, at the r.h.s. of the inequality. In such special problem the improvement of the estimates due to analyticity is clearly irrelevant.

all integrals are then extended[4] to $\text{Re}\, t_v \in {\rm I\!R}$, and (2.14) is used.

c) Finally, the sum over $\underline{\alpha}$ introduces in the estimate a factor $\prod_v (j_v + 1)$, less than $2^{h-1}$, while the sum over $\theta$ gives a factor $M_h$ = cardinality of $\Theta_h$; one easily recognizes that $M_h$ is precisely the quantity defined and estimated in Lemma 2, so one has in particular $M_h < 4^{h-1}$.

Putting everything together, one obtains

$$|\alpha_{h,\nu}| \leq W(8CW)^{h-1} \delta^{-hq+1} e^{-|\nu|\omega\tau(1-\delta)}, \qquad \alpha = \xi, \eta, \tag{3.15}$$

as claimed.

Item iv) of the proposition is a particular case of (3.13), (3.14).

### 3.3. The cancellation mechanism, and proof of point iii).

We show here that in the Hamiltonian case, due to a very special cancellation mechanism, the average energy $\eta_{h,0}$, which according to the above general estimate (3.15) does not contain any exponentially small factor, is instead proportional to $e^{-2\omega\tau(1-\delta)}$ (it is interesting to note that a similar property does not hold for $\xi_{h,0}$).

Let us consider again (3.13). The generic integral on $t_v$ ranges there from $-\infty$ to $t_{v'}$, $v'$ being the vertex which immediately precedes $v$ in the tree; the only exception is the integral associated with the root vertex (i.e. the vertex at the end of the root), which extends from $-\infty$ to $+\infty$. We then use the trivial identity

$$\int_{-\infty}^{t_{v'}} . \, dt_v = \frac{1}{2} \int_{-\infty}^{+\infty} . \, dt_v + \frac{1}{2} \int_{-\infty}^{+\infty} . \, S(t_{v'} - t_v)\, dt_v, \tag{3.16}$$

where $S$ is the sign function, and distinguish the two terms of this decomposition by a label $\lambda_v = 0$ for the former, $\lambda_v = 1$ for the latter. We then put a label $\lambda_v$, as a further decoration of the tree, on the branch leading to vertex $v$ (the root has necessarily $\lambda = 0$), and add in the formula for the value of the tree a primed sum over $\underline{\lambda}$ (a sum over all $v$ but the root); so we get

$$V(\theta, \underline{\alpha}, \underline{n}) = {\sum_{\underline{\lambda}}}' V(\theta, \underline{\alpha}, \underline{n}, \underline{\lambda}), \qquad \alpha_{h,\nu} = \sum_\theta {\sum_{\underline{\alpha}}}' \sum_{\underline{n}} {\sum_{\underline{\lambda}}}' V(\theta, \underline{\alpha}, \underline{n}, \underline{\lambda}), \tag{3.17}$$

with

$$V(\theta, \underline{\alpha}, \underline{n}, \underline{\lambda}) = \left(\prod_{v \in \theta} g_{n_v}\right) \int_{-\infty}^{\infty} \cdots \int_{-\infty}^{\infty} \prod_{v \in \theta} f_{\alpha_v}^{m_v, j_v}(t_v, \eta^o) \cdot \\ \cdot \frac{1}{2}[\lambda_v S(t_v - t_{v'}) + (1 - \lambda_v)]\, e^{i\omega \underline{n}\cdot \underline{t}}\, d\underline{t}. \tag{3.18}$$

It is very relevant to note that a label $\lambda_v = 0$ interrupts the chain of integrations, and the value of the tree factorizes, with one factor for each *cluster* of vertices internally connected by branches with $\lambda_v = 1$. For an intuitive picture we think that branches with $\lambda_v = 0$ are fragile and break, thus disconnecting the tree $\theta$ into $l \leq h - 1$ clusters $\mathcal{C}_0, \ldots, \mathcal{C}_{l-1}$; let $\mathcal{C}_0$ be the cluster connected to the root. As before, we can now raise the integration paths to $\text{Im}\, t_v = \pm\tau(1-\delta)$: the sign function $S$ is not analytic, but this does not cause any problem if the sign of $\text{Im}\, t_v$ is the same for all vertices of the same cluster; the sign of $\text{Im}\, t_v$ can instead be different for different clusters, and we can profit of this new freedom. Indeed, for each cluster $\mathcal{C}_i$ let us denote

$$\nu_i = \sum_{v \in \mathcal{C}_i} n_v \tag{3.19}$$

($\nu_0$ is called *free momentum* in [G1], [G2]) so that $\sum_{i=0}^{l-1} \nu_i = \nu = 0$; by appropriately choosing the sign of $\text{Im}\, t_v$, we then get

$$\left|e^{i\omega \underline{n}\cdot\underline{t}}\right| = e^{-N\omega\tau(1-\delta)}, \qquad N = \sum_{i=0}^{l-1} |\nu_i| \tag{3.20}$$

---

[4] Here the estimate is not optimal: by symmetrization of the integration domain, one could gain a combinatorial factor $\prod_v (j_v!)^{-1}$, but the final improvement on the constants $A$ and $B$ would be not so relevant.

Let us now introduce the decomposition

$$\alpha_{h,0} = \alpha'_{h,0} + \alpha''_{h,0} , \qquad \alpha = \xi, \eta , \tag{3.21}$$

where $\alpha'_{h,0}$ collects the contribution of decorated trees with $\nu_0 \neq 0$, while $\alpha''_{h,0}$ collects the contribution of trees with $\nu_0 = 0$. If $\nu = 0$ but $\nu_0 \neq 0$, then clearly $N \geq 2$. Proceeding just as we did above to prove (3.15), one immediately obtains

$$|\alpha'_{h,0}| \leq AB^{h-1}\delta^{-hq+1}e^{-2\omega\tau(1-\delta)} \tag{3.22}$$

(the only new fact is that each sum over $\lambda_v = 0, 1$ gives now an extra factor 2, which however is compensated by the factor $\frac{1}{2}$ in (3.18)).

We shall now restrict ourselves to $\alpha = \eta$, and use the Hamiltonian character of the problem. Our aim is to show that one has exactly $\eta''_{h,0} = 0$; in virtue of (3.22) this is clearly enough to prove point iii) of the proposition. To make evident the compensation, we must enumerate the trees differently. We now think of a tree with $h$ vertices $v_0, \ldots, v_{h-1}$ (before decorations are added) simply as a partially ordered set of $h$ symbols, and count exactly one tree for each "tree-like" partial ordering one can introduce in the set $\{0, \ldots, h-1\}$. Tree-like will mean that the following particular rules are satisfied:

i) one vertex (the root vertex) precedes all other vertices;

ii) if both $v'$ and $v''$ precede $v$, then either $v'$ precedes $v''$ or $v''$ precedes $v'$.

Note that a branch drawn from $v'$ to $v$ indicates that $v'$ immediately precedes $v$ (any other vertex preceeding $v$ also precedes $v'$ ).

The correspondence between old and new enumeration is seen after some meditation. For example, the three–vertices trees shown in figure 4a are now counted separately, while the two four–vertices trees exhibited in figure 4b are counted only once (they represent the same partially ordered set, simply drawn differently). The new set of trees with $h$ vertices will be called $\widetilde{\Theta}_h$. Concerning decoration, the only difference is that the labels $\alpha = \xi$ or $\eta$ on the internal branches are now completely free (there is no more top-to-bottom order); the $\underline{n}$ and $\underline{\lambda}$ decorations are left unchanged.

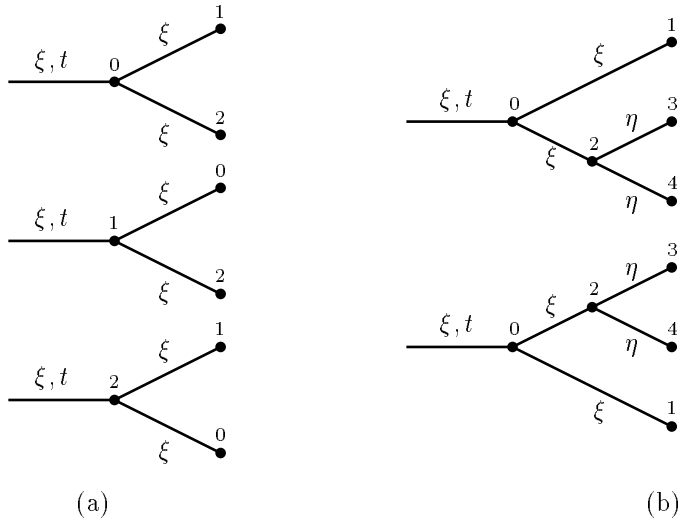

(a) (b)

*Figure 4.* Illustrating the new counting of trees as partially ordered sets: (a) different trees, to be counted separately; (b) different drawings of the same tree, to be counted only once.

As a result of the new way of counting (for a general reference on the combinatorial problems concerning various trees, see [HP]), one gets for the value of trees expressions almost identical to the previous ones, in particular to (3.17) and (3.18), with precisely the following differences: (a) the Taylor coefficients $f^{m_v,j_v}_{\alpha_v}$ loose the combinatorial factor $1/j_v!(m_v - j_v)!$, namely they are replaced by the derivatives

$$\frac{\partial^{m_v} f_{\alpha_v}}{\partial \xi^{j_v} \partial \eta^{m_v - j_v}} ; \tag{3.23}$$

(b) an overall external factor $1/(1+\sum_v m_v)! = 1/h!$ is added. So one has

$$\alpha_{h,\nu} = \frac{1}{h!} \sum_{\theta \in \widetilde{\Theta}} {\sum_{\underline{\alpha}}}' \sum_{\underline{n}} {\sum_{\underline{\lambda}}}' \widetilde{V}(\theta, \underline{\alpha}, \underline{n}, \underline{\lambda}) \qquad (3.24)$$

with

$$\widetilde{V}(\theta, \underline{\alpha}, \underline{n}, \underline{\lambda}) = \Big(\prod_{v \in \theta} g_{n_v}\Big) \int_{-\infty}^{\infty} \cdots \int_{-\infty}^{\infty} \prod_{v \in \theta} \frac{\partial^{m_v} f_{\alpha_v}}{\partial \xi^{j_v} \partial \eta^{m_v - j_v}}(t_v, \eta^o) \\ \frac{1}{2}[\lambda_v S(t_v - t_{v'}) + (1 - \lambda_v)] \, e^{i\omega \underline{n} \cdot \underline{t}} \, d\underline{t} \qquad (3.25)$$

We now focus the attention on the set of trees contributing to $\eta''_{h,0}$, and divide it into equivalence classes; our purpose will be to show that one has exact compensation within each class. To define classes, we introduce the operation of "moving the root" of a tree (the idea is taken from a result of Chierchia taken from [G1]: see the cancellation mechanism IV in §6 of [G1] and the related lemma). First of all, let us note that in any tree the path from the root vertex, say $v_0$, to any other vertex, say $v_1$, is uniquely defined. Moving the root from $v_0$ to $v_1$ simply means modifying the order relation (i.e., replacing the tree inside $\widetilde{\Theta}$ by a different one) just by inverting the orientation of branches along the path. The idea, actually very intuitive, is illustrated in figure 5 (the change in the decoration will be explained in a moment). Note that the operation is well defined since, for a given set of (non oriented) branches, the choice of the root vertex (of the first element of the tree–like partially ordered set) uniquely determines the orientation (the partial ordering).

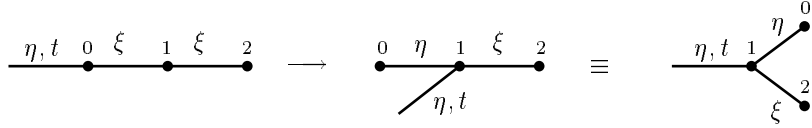

*Figure 5.* Illustrating the operation of moving the root in a tree. The root moves along the branch joining vertices 0 and 1; correspondingly the branch changes orientation, and its label $\xi$ turns into $\eta$. The second and the third graphs are simply different drawings of the same tree.

We now say that two decorated trees $(\theta, \underline{\alpha}, \underline{n}, \underline{\lambda})$ and $(\theta', \underline{\alpha}', \underline{n}', \underline{\lambda}')$ are equivalent, if:

a) $\theta'$ is obtained from $\theta$ by moving the root inside the root cluster $\mathcal{C}_0$.
b) The decoration $\underline{\alpha}'$ differs from $\underline{\alpha}$ along the path: namely any label $\xi$ is there converted into $\eta$, and conversely (see figure 5); in compact notation, say that the label $\alpha_b$ of branch $b$ in the path is converted into its Hamiltonian conjugated, denoted $\overline{\alpha}_b$.
c) All other decorations are left unchanged, so $\underline{n} = \underline{n}'$ and $\underline{\lambda} = \underline{\lambda}'$.

To illustrate the compensation mechanism, it is convenient to introduce the following notation:

$B_0 = $ the set of branches of $\mathcal{C}_0$
$v'_b, v''_b = $ the ends (tail and tip respectively) of branch $b \in B_0$.
$k_0 = $ the number of $\eta$–type branches in $B_0$.
$\underline{t}_0, \underline{n}_0 = $ the restrictions of $\underline{t}$ and $\underline{n}$ to the cluster $\mathcal{C}_0$.
$\partial_\alpha^{(v)} = $ the derivative with respect to $\alpha = \xi$ or $\eta$, which however operates only on $f^{(v)}$.

We can write the value $\widetilde{V}$ of a decorated tree in the factorized form

$$\widetilde{V}(\theta, \underline{\alpha}, \underline{n}, \underline{\lambda}) = V^{(0)}(\theta, \underline{\alpha}, \underline{n}) \, V^{(1)}(\theta, \underline{\alpha}, \underline{n}, \underline{\lambda}) \qquad (3.26)$$

where $V^{(0)}$ is the contribution of the cluster $\mathcal{C}_0$, while $V^{(1)}$ collects the contributions of the remaining clusters; $V^{(1)}$ is clearly constant within the equivalence class. Using (finally!) the Hamiltonian character of the problem, namely definition (2.9) for $f_\xi$ and $f_\eta$, one recognizes that for each branch $b$ one has exactly one pair of derivatives $\partial_{\alpha_b}^{(v'_b)}$ and $\pm\partial_{\overline{\alpha}_b}^{(v''_b)}$, with the minus sign if $\alpha_b = \eta$; precisely, one has

$$V^{(0)}(\theta, \underline{\alpha}, \underline{n}) = \Big(\prod_{v \in \mathcal{C}_0} g_{n_v}\Big) \int_{-\infty}^{+\infty} \cdots \int_{-\infty}^{+\infty} U(\theta, \underline{\alpha}, \underline{n}; \underline{t}_0) \, e^{i\omega \underline{n} \cdot \underline{t}} \, d\underline{t}_0 \qquad (3.27)$$

with

$$U(\theta, \underline{\alpha}, \underline{n}; \underline{t}_0) = (-1)^{k_0} \Big[ \prod_{b \in B_0} \frac{1}{2} S(t_{v'_b} - t_{v''_b}) \Big] \partial_\xi^{(v_0)} \prod_{b \in B_0} \partial_{\alpha_b}^{(v'_b)} \partial_{\overline{\alpha}_b}^{(v''_b)} \prod_{v \in \mathcal{C}_0} f^{(v)}(t_v, \eta^o) \qquad (3.28)$$

where $f^{(v)}$ is $f$, or a convenient derivative of $f$ in case there are branches with $\lambda = 0$ coming out of $v$. Let us stress that $\partial_\alpha^{(v)}$ acts only on $f^{(v)}$. The outermost derivative $\partial_\xi^{(v_0)}$ refers to the root vertex, here denoted $v_0$ (accordingly to (2.9), the derivative is with respect to $\xi$, since the label of the root is here $\eta$).

We can now see that

$$U(\theta, \underline{\alpha}, \underline{n}; \underline{t}_0) = \partial_\xi^{(v_0)} \mathcal{U}(\theta, \underline{\alpha}, \underline{n}; \underline{t}_0) \;, \qquad (3.29)$$

with $\mathcal{U}$ independent of the tree within the equivalence class. Indeed, whenever one moves the root from one vertex to a nearby one (it is clearly enough to consider this case), a branch $b$ of $\mathcal{C}_0$ is inverted: correspondingly $v'_b$ e $v''_b$ (tail and tip of $b$) interchange, but on the same time $\alpha_b$ is replaced by $\overline{\alpha}_b$ (and conversely), so the product $\partial_{\alpha_b}^{(v'_b)} \partial_{\overline{\alpha}_b}^{(v''_b)}$ is left invariant. The sign function $S$ relative to the inverted branch changes its sign, but the number $k_0$ of $\eta$–type branches changes exactly by one unity, so $\mathcal{U}$ does not change.

The conclusion now follows: by taking the sum over all trees in the same equivalence class, one clearly gets

$$\Big( \sum_{v \in \mathcal{C}_0} \partial_\xi^{(v)} \Big) \mathcal{U}(\theta, \underline{\alpha}, \underline{n}; \underline{t}_0) = \Big( \frac{\mathrm{d}}{\mathrm{d} t_1} + \cdots + \frac{\mathrm{d}}{\mathrm{d} t_{h_0}} \Big) \mathcal{U}(\theta, \underline{\alpha}, \underline{n}; \underline{t}_0) \qquad (3.30)$$

where each derivative $\frac{\mathrm{d}}{\mathrm{d} t_v}$ is intended to act only on $f^{(v)}$ inside $\mathcal{U}$, and not on sign functions. One can finally integrate by parts: for the assumptions on $f$ at infinity the finite term vanishes; the derivatives of the sign functions give no contribution (more precisely, one easily sees that each of them gives two opposite contributions); and when the derivatives are transferred to the exponential, one gets in front an overall factor $\sum_{v \in \mathcal{C}_0} n_v = \nu_0 = 0$. This proves the exact compensation inside a class, so that $\eta''_{h,0} = 0$, and concludes the proof of the Proposition.

## 4. Concluding remarks.

### 4.1. I–dependent, non factorized perturbations.

The Hamiltonian (2.1) that we have treated, has a rather special form; in particular, the coupling term $\varepsilon g(\varphi) \widetilde{V}_1(x)$ is factorized and independent of $I$. Such Hamiltonian, properly speaking, does not represent the motion of an oscillator in an external potential, rather a point mass which is periodically perturbed with fixed frequency $\omega$. The consequence of our choice was that, once the coordinates $\xi$, $\eta$ were introduced, the equations of motion assumed the special form (2.8), namely they reduced to only one pair of (time dependent) equations for $\xi$ and $\eta$. For this reason we could write a perturbative scheme to compute $\xi(t)$ and $\eta(t)$ at all orders in $\varepsilon$, treating separately (and trivially) the motion of $\varphi$, and completely ignoring $I$.

However, the perturbative scheme we developed adapts very easily to Hamiltonians of the form

$$K(\xi, \eta, \varphi, I) = \omega I + \eta + \varepsilon f(\xi, \eta, \varphi, I) \;, \qquad (4.1)$$

or more generally, to equations of motion of the form

$$\begin{aligned} \dot\xi &= 1 + \varepsilon f_\xi(\xi, \eta, \varphi, I) & \dot\varphi &= \omega + \varepsilon f_\varphi(\xi, \eta, \varphi, I) \\ \dot\eta &= \varepsilon f_\eta(\xi, \eta, \varphi, I) & \dot I &= \varepsilon f_I(\xi, \eta, \varphi, I) \end{aligned} \qquad (4.2)$$

with, in the Hamiltonian case, $f_\xi = \frac{\partial f}{\partial \eta}$, $f_\eta = -\frac{\partial f}{\partial \xi}$, $f_\varphi = \frac{\partial f}{\partial I}$, $f_I = -\frac{\partial f}{\partial \varphi}$.

We indicate briefly how to proceed in this more general situation, and to this end, we introduce the following compact notation:

$\underline{X} = (\xi, \eta, \varphi, I)$

$\alpha = $ a label running on the set $\mathcal{A} = \{\xi, \eta, \varphi, I\}$

$$\underline{j} = (j_\xi, j_\eta, j_\varphi, j_I) \in \mathbb{N}^4; \quad |\underline{j}| = \sum_\alpha j_\alpha; \text{ and } \underline{j}! = j_\xi! \cdots j_I!$$

$$\underline{\underline{k}} = \{k_i^\alpha \in \mathbb{N}_+ : \alpha \in \mathcal{A}, \, 0 \le i \le j_\alpha\}, \text{ and } |\underline{\underline{k}}| = \sum_{\alpha,i} k_i^\alpha$$

$$\underline{X}(t) = \underline{X}_0(t) + \sum_h \varepsilon^h \underline{X}_h(t)$$

$$\underline{X}^{\underline{j}} = \xi^{j_\xi} \cdots I^{j_I}$$

$$f_\alpha^{\underline{j}} = \frac{1}{\underline{j}!} \frac{\partial^{|\underline{j}|} f_\alpha}{\partial \underline{X}^{\underline{j}}}.$$

The recursive scheme that one gets, for a perturbative representation of the motions satisfying the asymptotic conditions

$$\xi(t) - t \to 0, \qquad \eta(t) \to \eta^\circ, \qquad \varphi(t) - \omega t \to \varphi^\circ, \qquad I(t) \to I^\circ \tag{4.3}$$

for $t \to -\infty$, is

$$\begin{aligned}
\alpha_h(t) &= \int_{-\infty}^t F_h^\alpha(t') \, dt' \\
F_h^\alpha(t) &= \sum_{\underline{j}: 1 \le |\underline{j}| \le h-1} f_\alpha^{\underline{j}}(X_0(t)) \sum_{\underline{\underline{k}}: |\underline{\underline{k}}| = h-1} \prod_{1 \le i_\xi \le j_\xi} \cdots \prod_{1 \le i_I \le j_I} \xi_{k_{i_\xi}^\xi} \cdots I_{k_{i_I}^I}
\end{aligned} \tag{4.4}$$

Note that $|\underline{j}|$ replaces here $m$.

Such a scheme gives rise to tree expansions quite similar to the previous ones; the only difference is that now the labels $\alpha$ on root and branches run over the whole set $\mathcal{A}$, and not only on $\xi$ and $\eta$. One then introduces, as before, the Fourier expansion of the perturbation, here

$$f_\alpha^{\underline{j}}(\xi, \eta, \varphi, I) = \sum_{\nu \in \mathbb{Z}} \mathcal{F}_{\alpha,\nu}^{\underline{j}}(\xi, \eta, I) \, e^{i\nu\varphi} \tag{4.5}$$

and collects terms which contribute to the same Fourier component $\alpha_{h,\nu}$, $\alpha \in \mathcal{A}$; this means introducing sums in the perturbative expansion, and adding decorations to the tree, exactly as before. The estimates then run quite smoothly, and one finds exponential estimates, i.e.:

$$|\alpha_{h,\nu}| \le AB^{h-1} \delta^{-hq+1} e^{-|\nu|\omega\tau(1-\delta)} \tag{4.6}$$

precisely as before. An assumption generalizing (2.10), (2.13) is

$$\left| \mathcal{F}_{\alpha,\nu}^{\underline{j}}(s + i\sigma, \eta^\circ, I^\circ) \right| \le C^{|\underline{j}|} \delta^{-|\underline{j}|-m_0} \, w(s) \, e^{-\rho|\nu|} \qquad \text{for } |\sigma| \le \tau(1-\delta) \tag{4.7}$$

In the Hamiltonian case, and only for $\alpha = \eta$ or $\alpha = I$ (as is obvious, $\Delta\eta$ and $\Delta I$ are not independent, since for energy conservation one has asymptotically $\Delta\eta = -\omega\Delta I$), one has then the cancellation mechanism, leading to the exponential laws

$$|\alpha_{h,0}| \le AB^{h-1} \delta^{-hq+1} e^{-2\omega\tau(1-\delta)}, \qquad \alpha = \eta, I \tag{4.8}$$

The method is exactly the same: one uses the identity (2.23), and introduces clusters of vertices as before; everything repeats then precisely as in the previous case, with the only obvious generalization that, when the root is moved along a path in the cluster $\mathcal{C}_0$, labels $\varphi$ and $I$ also interchange. The conclusions, essentially expression (3.17) for $\eta$ and a very similar one for $I$, follow. In a word: the case of Hamiltonian (4.1) is practically identical to the simpler case we treated in detail; only the notation gets complicated.

### 4.2. Reversible systems.

The cancellation mechanism leading to the special exponential law for $\eta_{h,0}$ was based in an essential way on the Hamiltonian character of the equations of motion. It is worthwhile to remark that an alternative assumption leading to the same result is a symmetry in the system, namely that $f_\xi(\xi, \eta)$ and $f_\eta(\xi, \eta)$ are respectively even and odd in $\xi$, and $g(\varphi)$ is even in $\varphi$. The time reversal symmetry of (1.1), i.e. the transformation $i : (p_x, p_y, x, y) \to (-p_x, -p_y, x, y)$, becomes in the $\xi, \eta, \varphi, I$ variables $i' : (\xi, \eta, \varphi, I) \to (-\xi, \eta, -\varphi, I)$: hence the above parity properties imply time reversal invariance of the equations of motion (*i.e.* they imply that $i'S_t = S_{-t}i'$, if $t \to S_t$ is the flow generated by the equations of motion). Hence reversibility can replace the Hamiltonian character.

In such a non–Hamiltonian but reversible cases the proof that $\eta''_{h,0} = 0$ proceeds as follows. First of all, it is not necessary to change the enumeration of trees (i.e., redefining trees as ordered sets and so on). Each tree can be described as a *core*, namely the root cluster $\mathcal{C}_0$ fully composed of branches with $\lambda = 1$, carrying several (possibly none) "sub–trees" connected to the core by branches with $\lambda = 0$ (the roots of the sub–trees). Instead of moving the root, here we move sub–trees, more precisely sub–trees with $\xi$-type root, if any. Let us sketch the procedure. Assume for a moment that at least one $\xi$-type sub–tree exists, and has multiplicity one, namely it is different from all other sub–trees, and consider the "pruned tree" $\widehat{\theta}$ one obtains by just removing this sub–tree. The value $V(\widehat{\theta}, \underline{\alpha}, \underline{n}, \underline{\lambda})$ is obtained by integrating the functions (see (3.18)):

$$\prod_{v \in \widehat{\theta}} f^{m_v, j_v}_{\alpha_v}(t_v, \eta^o) \frac{1}{2}[\lambda_v S(t_v - t_{v'}) + (1 - \lambda_v)] e^{i\omega \underline{n} \cdot \underline{t}} . \tag{4.9}$$

By adding the sub–tree to any vertex $\widehat{v}$ of $\mathcal{C}_0$, in all $j_{\widehat{v}} + 1$ possible ways, one gets

$$\left[\prod_{v \neq \widehat{v}} f^{m_v, j_v}_{\alpha_v}(t_v, \eta^o)\right] (j_{\widehat{v}} + 1) f^{m_{\widehat{v}} + 1, j_{\widehat{v}} + 1}_{\alpha_{\widehat{v}}}(t_{\widehat{v}}, \eta^o) \prod_{v \in \widehat{\theta}} \frac{1}{2}[\lambda_v S(t_v - t_{v'}) + (1 - \lambda_v)] e^{i\omega \underline{n} \cdot \underline{t}} \tag{4.10}$$

But clearly, one has

$$(j_{\widehat{v}} + 1) f^{m_{\widehat{v}} + 1, j_{\widehat{v}} + 1}_{\alpha_{\widehat{v}}} = \frac{\partial}{\partial \xi} f^{m_{\widehat{v}}, j_{\widehat{v}}}_{\alpha_{\widehat{v}}} , \tag{4.11}$$

so that, also taking the sum for $\widehat{v} \in \mathcal{C}_0$, one gets

$$\left(\frac{\mathrm{d}}{\mathrm{d}t_1} + \ldots + \frac{\mathrm{d}}{\mathrm{d}t_{h_0}}\right) \prod_{v \in \theta} f^{m_v, j_v}_{\alpha_v}(t_v, \eta^o) \frac{1}{2}[\lambda_v S(t_v - t_{v'}) + (1 - \lambda_v)] e^{i\omega \underline{n} \cdot \underline{t}} , \tag{4.12}$$

where the derivative $\frac{\mathrm{d}}{\mathrm{d}t_v}$ acts only on $f^{m_v, j_v}_{\alpha_v}$ (and not on sign functions containing $t_{\widehat{v}}$). Integrating by parts as before, one easily gets $\alpha''_{h,0} = 0$, $\alpha = \xi, \eta$.

In the case there are $\xi$-type subtrees, but all of them have multiplicity greater than one, one can still move any of them, say of multiplicity $k$; simple combinatorial arguments then show that one gets

$$\left(\frac{\mathrm{d}}{\mathrm{d}t_1} + \ldots + \frac{\mathrm{d}}{\mathrm{d}t_{h_0}}\right)^k \tag{4.13}$$

in place of $\frac{\mathrm{d}}{\mathrm{d}t_1} + \ldots + \frac{\mathrm{d}}{\mathrm{d}t_{h_0}}$ in the above expression, and the conclusion does not change.

The operation of "moving a sub–tree", here introduced, closely resembles the operation of "moving a fruit" in [G1] (see mechanism II in §6 of [G1]).

So far the symmetry of $f$ and $g$ has not been exploited;[5] we use it now, to take care of trees without any $\xi$-type subtree. One easily checks that, *if the root is of $\eta$-type*, then the quantity to be integrated is exactly antisymmetric for the exchange $\underline{t}_0 \to -\underline{t}_0$, and simultaneously $\underline{n}_0 \to -\underline{n}_0$, in the root cluster. Indeed:

i) The number of $S$ functions which change sign is $h_0 - 1$.

ii) The number of derivatives with respect to $\xi$, which obviously change the parity of $f_\alpha$ functions, is $\sum_{v \in \mathcal{C}_0} j_v$.

iii) The number of odd functions $f_\eta$ (including the one associated with the root) is $1 + \sum_{v \in \mathcal{C}_0} (m_v - j_v)$.

The overall parity is thus $h_0 + \sum_{v \in \mathcal{C}_0} m_v = 2h_0 - 1$, namely odd. On the other hand, the scalar product $\underline{n} \cdot \underline{t}$ is invariant, while, by the assumed parity of $g$, one has $g_n = g_{-n}$. Since all integrals run from $-\infty$ to $\infty$, one concludes that trees with opposite decorations $\underline{n}$ and $-\underline{n}$ have opposite value, thus they exactly compensate when one computes $\eta''_{h,0}$. The latter cancellation mechanism is close to the "parity cancellation" (see mechanism I in §6 of [G1]).

In essence: the above analysis puts in evidence the terms which are possibly large, among those contributing to $\Delta \eta$, namely those associated to trees with $\nu = 0$, $\nu_0 = 0$, and no $\xi$-like branches issuing from $\mathcal{C}_0$. Such terms cannot be proven to be small on the basis of purely perturbative arguments (they would not be small, for example, for a problem with friction): they compensate exactly only in presence of a symmetry, namely either the Hamiltonian character or the reversibility of the equations of motion.

---

[5] Hence the above argument has general validity, and it could also be used to (slightly) improve constants in the estimates of Section 3.

### 4.3. The problem of the energy exchange as a problem of split separatrices.

Here we provide a (rather informal) geometric picture, which can be useful to interpret our results; in particular, we establish the link between the problem of the energy exchange, which was discussed in the previous sections, and the idea of separatrices splitting.

To this purpose, let us consider again the Hamiltonian $K$ given in (2.6), and look at the three–dimensional energy surface $\Gamma_E$ corresponding to some fixed value of the total energy $E$. Good coordinates on $\Gamma_E$ are $(\xi, \eta, \varphi)$, but for an easier insight it is better to look at $\varphi^o = \varphi - \omega t$ in place of $\varphi$. We fix the attention to the set of trajectories in $\Gamma_E$, such that $\xi(t) - t \to 0$ asymptotically for $t \to -\infty$.[6] One has then one trajectory (one curve in $\Gamma_E$) for each choice of $\eta^o = \lim_{t \to -\infty} \eta(t)$ and of $\varphi^o$. By varying $\varphi^o$ at fixed $\eta^o$, one gets a two–dimensional regular submanifold $\Sigma^-_{\eta^o, \varepsilon}$ of $\Gamma_E$ (in fact, an analytic surface), whose equation can be written in the form $\eta = \eta^-_{\eta^o, \varepsilon}(\xi, \varphi^o)$. Clearly $\Sigma^-_{\eta^o, \varepsilon}$ coincides, for $\varepsilon = 0$, with the plane (the cylinder, if one recalls that $\varphi$ is an angle) of constant $\eta = \eta^o$. For $\varepsilon \neq 0$, $\Sigma^-_{\eta^o, \varepsilon}$ is (by definition) flat asymptotically for $\xi \to -\infty$, while, according to Proposition 1 (Section 2), it assumes a sinusoidal shape, of exponentially small amplitude, asymptotically for $\xi \to +\infty$; in the middle, namely in correspondence with the collision, the oscillations are instead much larger, namely of order $\varepsilon$. The behavior of $\Sigma^-_{\eta^o, \varepsilon}$ is represented in figure 6 (gray surface).

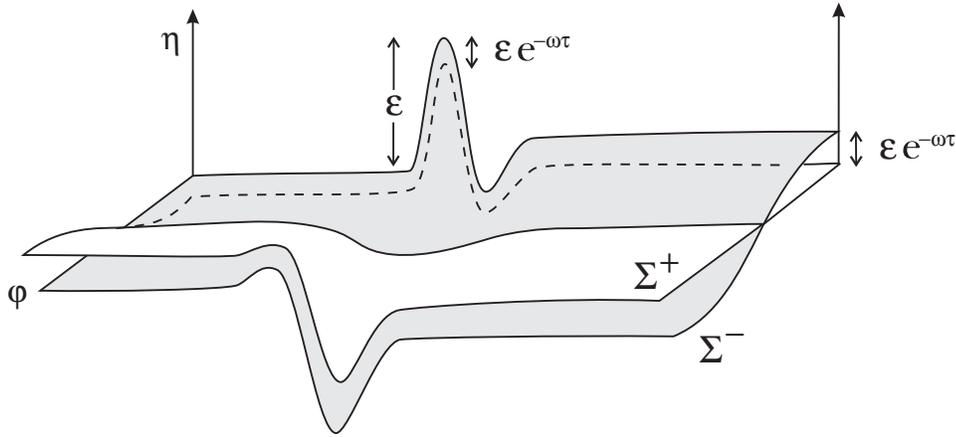

*Figure 6.* Illustrating the behavior of $\Sigma^+_{\eta^o, \varepsilon}$ (white) and $\Sigma^-_{\eta^o, \varepsilon}$ (grey).

Exactly in the same way, one can look at trajectories such that $\eta \to \eta^o$ in the opposite limit $t \to +\infty$. The corresponding curves form a second surface $\Sigma^+_{\eta^o, \varepsilon}$, described by an equation of the form $\eta = \eta^+_{\eta^o, \varepsilon}(\xi, \varphi^o)$, which is instead flat for $\xi \to +\infty$, and assumes a sinusoidal shape (of exponentially small amplitude) for $\xi \to -\infty$, with oscillations of amplitude $\mathcal{O}(\varepsilon)$ in the middle (figure 6, white surface). We claim that the two surfaces intersect each other as is symbolically shown in figure 6, more precisely:

i. By taking a section $\varphi^o = \text{const.}$, one gets two curves $\sigma^+_{\eta^o, \varepsilon}(\varphi^o)$ and $\sigma^-_{\eta^o, \varepsilon}(\varphi^o)$ which either coincide (this happens for those special values of $\varphi^o$, such that the asymptotic energy exchange $\Delta \eta$ vanishes; there are two of them for any $\eta^o$), or never intersect.

Indeed, should they intersect once, they would intersect infinitely many times, and this is in conflict with the existence of well defined limits of both curves in both directions.

ii. Although $\eta^+_{\eta^o, \varepsilon}$ and $\eta^-_{\eta^o, \varepsilon}$ are in a sense large (namely of order $\varepsilon$) during the interaction, their difference at fixed $\xi$ and $\varphi^o$ is always exponentially small:

$$\eta^+_{\eta^o, \varepsilon}(\xi, \varphi^o) - \eta^-_{\eta^o, \varepsilon}(\xi, \varphi^o) = \mathcal{O}(e^{-\tau \omega}) \tag{4.14}$$

so, in particular, a generic section $\varphi^o = \text{const.}$ looks like figure 7.

---

[6] The choice is made for coherence with the previous notations; actually, one could equally well make the more symmetric choice $\xi(0) = 0$.

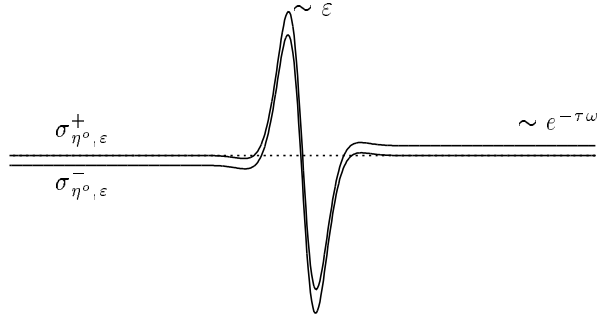

*Figure 7.* Possible profiles of $\sigma^+_{\eta^o,\varepsilon}$ and $\sigma^-_{\eta^o,\varepsilon}$.

This follows from volume preservation of Hamiltonian dynamics: indeed, in place of $K$, one can equivalently consider, for the evolution of $\xi$ and $\eta$, the non–autonomous Hamiltonian

$$\widetilde{K}(\xi,\eta,t;\varphi^o) = \eta + \varepsilon g(\varphi^o + \omega t) f(\xi,\eta) \tag{4.15}$$

depending parametrically on $\varphi^o$. The area $d\xi d\eta$ is clearly preserved by the Hamiltonian dynamics; on the other hand, $d\xi$ is almost preserved, as follows from $d\xi = \dot\xi\, dt = (1 + \mathcal{O}(\varepsilon))\, dt$. So $d\eta$ is almost preserved, too, and any vertical segment in figure 6 — in particular, the vertical separation of $\sigma^\pm_{\eta^o,\varepsilon}(\varphi^o)$ at any $\xi$ — is in turn preserved by the dynamics, up to a factor $1 + \mathcal{O}(\varepsilon)$.

iii. In any section $\xi = $ const. one obtains split curves $\tilde\sigma^\pm_{\eta^o,\varepsilon}$, which intersect transversally, but for any $\xi$ their separation is exponentially small (see figure 8).

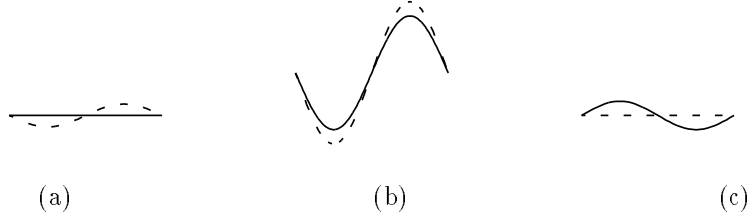

(a)          (b)          (c)

*Figure 8.* Possible profiles of $\tilde\sigma^-_{\eta^o,\varepsilon}$ (solid line) and $\tilde\sigma^+_{\eta^o,\varepsilon}$ (dashed line) at $\xi = -\infty$ (a), during the collision (b), and at $\xi = +\infty$ (c).

This is an obvious consequence of the previous analysis, which however shows an important point: the problem of the energy exchange in our system can be thought of as a problem of separatrices splitting. This is relevant, in our opinion, for the connection with [G1,G3] (as a matter of fact, this was our first approach to the problem: only later we recognized that the direct computation of the energy exchange is simpler).

The special form of the perturbation in the Hamiltonian (2.6) is not relevant, and essentially identical conclusions can be drawn for the more general case considered in (4.1).

### 4.4. Systems of several oscillators.

An interesting question is whether the above analysis and results can be extended to systems with more than one fast degrees of freedom. Work is in progress in this direction. We anticipate here just one result, concerning a natural extension of Hamiltonian (2.6), namely

$$K(\xi,\eta,\underline{\varphi},\underline{I}) = \underline{\omega} \cdot \underline{I} + \eta + \varepsilon f(\xi,\eta) g(\underline{\varphi}) , \tag{4.16}$$

with $\xi, \eta, \in \mathbb{R}$, $\underline{\varphi} = (\varphi_1,\ldots,\varphi_l) \in \mathbb{T}^l$, $\underline{I} = (I_1,\ldots,I_l) \in \mathbb{R}^l$. Such a system represents a point mass which is perturbed in a quasi–periodic way, with $l$ frequencies.

We keep the assumptions (2.13), (2.14) for $f$, and trivially generalize (2.10) to $|g(\underline{\varphi})| \leq 1$ for $|\mathrm{Im}\,\varphi_j| < \rho$, $j = 1,\ldots,l$. Proposition 1 can be extended as follows:

**Proposition 2.** *Consider the Hamiltonian (4.16); let $f$ and $g$ satisfy the above assumptions, and consider motions satisfying the asymptotic conditions (2.15). Then:*

i) *For small $\varepsilon$ the Taylor series (2.17) converge, precisely for any real $t$ one has*

$$|\alpha_h(t)| \leq A B^{h-1} , \qquad \alpha = \xi, \eta \tag{4.17}$$

*ii)* For any $\underline{\nu} \in \mathbb{Z}^l$, the Fourier component $\alpha_{h,\underline{\nu}}$ of $\alpha_h(+\infty)$, $\alpha = \xi, \eta$, satisfies the exponential estimate

$$|\alpha_{h,\underline{\nu}}| \leq AB^{h-1}\delta^{-hq+1}e^{-\tau|\underline{\nu}\cdot\underline{\omega}|(1-\delta)-\rho|\underline{\nu}|} \ . \tag{4.18}$$

*iii)* For $\underline{\nu} \in \mathbb{Z}^l$ such that $\underline{\nu} \cdot \underline{\omega} = 0$, one has

$$|\eta_{h,\underline{\nu}}| \leq AB^{h-1}\delta^{-hq+1}e^{-2\tau|\underline{\nu}'\cdot\underline{\omega}|(1-\delta)-2\rho|\underline{\nu}'|} \ , \tag{4.19}$$

with some $\underline{\nu}'$ such that $\underline{\nu}' \cdot \underline{\omega} \neq 0$.

*iv)* At first order in $\varepsilon$ one has the explicit expression

$$\eta_{1,\underline{\nu}} = i(\underline{\nu} \cdot \underline{\omega})\, g_{\underline{\nu}} \int_{-\infty}^{\infty} f(t, \eta^o)\, e^{i|\underline{\nu}\cdot\underline{\omega}|t}\, \mathrm{d}t \ . \tag{4.20}$$

Since $\underline{\nu} \cdot \underline{\omega}$ can be arbitrarily small for large $\nu$, the therm $-\rho|\nu|$ at exponent in (4.18) is now very important. The exponential estimate (4.19) is the analog of (2.22), and is based on cancellations. Expression (4.20) represents the JLT approximation in this problem. Estimates of the form (4.18) are also found in [G1] (see (8.1)) for separatrix splitting problems (they were, however, derived with a different technique relaying on the classical proofs of the KAM theorem), and in [Ge], where they are derived by using a tree representation of the $\alpha_{h,\nu}$. See also [Si], where a first order analysis is performed in connection with the related problem of separatrices splitting.

For a detailed analysis of the JLT approximation in systems with several fast variables, including a careful comparison with numerical results (quite nice agreement is found), see [BCF].

### 4.5. Improving the bounds.

The presence of the factor $(1-\delta)$ in the exponential bounds (2.21) is clearly a sign that some of the estimates can be improved. Of course one needs more information about the nature of the singularity at $\mathrm{Im}\, t = \pm i$. If we keep the assumption that $V_0 = e^{-x}$ and $V_1 = e^{-x}$ then we see that $f_\xi, f_\eta$ have a pole of order $m_0 = 3$ in the $t$ variable at $\pm i\tau$.

This means that the integrals on the $t_v$ variables can be shifted to larger imaginary parts than $\tau$, leaving, besides the integrals at $\mathrm{Im}\, t_v = \tau(1+\delta)$, a few integrals over segments $\pm[i\tau, i\tau(1+\delta)]$ (due to logarithmic singularities generated from the poles by the integrations) plus some residues from the singularities at $\pm i\tau$. Since there are $h$ integrals the residues, which provide the most singular contributions, will contribute quantities bounded by:

$$(C\sum_v |n_v|\omega)^{(m_0-1)h} e^{-|\nu|\omega\tau} \tag{4.21}$$

Hence if $g(\varphi)$ is a trigonometric polynomial of degree $N$ our estimates can be improved, with some extra work, to:

$$h^h (\omega\, NC)^{2h} e^{-\omega\tau\, (|\nu|+2\delta_{0,\nu})} \tag{4.22}$$

This shows, by combining the latter bound with the previous and by optimizing the choices, that in the above cases the condition for having domination of the lowest order is essentially (up to logarithms) $\varepsilon\omega^3 \ll 1$ (instead of the $\varepsilon\omega^4 \ll 1$) discussed in §1.

*Acknowledgements:* we are indebted to F. Fassò, L. Galgani, G. Gentile and J. Strelcyn for useful discussions. This work has been developed with the support of the grant "EC contract ERBCHRXCT940460" for the project "Stability and universality in classical mechanics".

**The Appendix**

As already remarked in the Introduction, the use of the $\xi, \eta$ variables simplifies the perturbative scheme, but is not strictly necessary. We give here a few indications on how to proceed with the original $p, q$ variables; this may be useful on the one hand for a comparison with [G1] and [G2], on the other hand because, although the scheme gets more complicated for the presence of nontrivial Wronskians and kernels in integral expressions, some points, like the analyticity requirements on the perturbing function, are perhaps more immediately understood.

The system we aim to study is Hamiltonian (2.1), with exponential potentials. However, to make more transparent the perturbative construction, and emphasize in particular the analyticity assumptions we need, we write the equations of motion in the more general form

$$\dot{x} = p, \qquad \dot{p} = \Phi(x) + \varepsilon g(\varphi) f(x), \qquad \dot{\varphi} = \omega, \qquad \dot{I} = \gamma(\varphi) V(x) ;$$

in the problem at hand one has clearly $\Phi(x) = f(x) = V(x) = e^{-x}$, and $\gamma(x) = -g'(\varphi)$. We look for motions such that, asymptotically for $t \to -\infty$,

$$p(t) \to p(-\infty) \equiv -\sqrt{2\eta}, \quad x(t) - p(-\infty)t \to 0, \quad \varphi(t) - \omega t \to \varphi^o, \quad I(t) \to I(-\infty) \equiv I^o ;$$

we aim to compute $\Delta I = I(+\infty) - I(-\infty)$.

For $\varepsilon = 0$, and $\Phi(x) = e^{-x}$, one has the unperturbed motion (see (2.5))

$$x_0(t) = 2\log[\eta^{-1}(\cosh\sqrt{\eta/2}\,t)^2], \quad p_0(t) = \sqrt{2\eta}\tanh\sqrt{\eta/2}\,t, \quad \varphi_0(t) = \varphi^o + \omega t, \quad I_0(t) = I^o ,$$

and the associated Wronskian, restricted to the first two components, is

$$\mathcal{W}(t) = \begin{pmatrix} w_{xx}(t) & w_{xp}(t) \\ w_{px}(t) & w_{pp}(t) \end{pmatrix} = \begin{pmatrix} 1 - \sqrt{\frac{\eta}{2}}\,t\tanh\sqrt{\frac{\eta}{2}}\,t & \sqrt{\frac{2}{\eta}}\tanh\sqrt{\frac{\eta}{2}}\,t \\ -\tanh\sqrt{\frac{\eta}{2}}\,t - \frac{\eta}{2}[\cosh\sqrt{\frac{\eta}{2}}\,t]^{-2}t & [\cosh\sqrt{\frac{\eta}{2}}\,t]^{-2} \end{pmatrix} .$$

For $\varepsilon \neq 0$ we write

$$x(t) = x_0(t) + \sum_{h>0} \varepsilon^h x_h(t), \qquad p(t) = p_0(t) + \sum_{h>0} \varepsilon^h p_h(t), \qquad I(t) = I^o + \sum_{h>0} \varepsilon^h I_h(t) ,$$

and all series will converge provided, in the series for $x(t)$, one extracts at each order $h$ a common factor $\beta(t)$ growing linearly with $|t|$: say, to fix the ideas, $\beta(t) = \sqrt{1+t^2}$ (such a linear growth is due to the element $w_{xx}$ in the wronskian above).

We then expand $\Phi$, $f$ and $V$ in Taylor series around the unperturbed motion $x_0(t)$:

$$\Phi(x(t)) = \Phi(x_0(t)) + \sum_{m\geq 1} \Phi^m(x_0(t))[x(t) - x_0(t)]^m , \qquad \text{with} \quad \Phi^m = \frac{1}{m!}\frac{\partial^m \Phi}{\partial x^m} ,$$

and similarly for $f$ and $V$. One must require that such series converge, although, as remarked above, $x(t) - x_0(t)$ diverges linearly in $t$. The simplest situation is that of integer functions, like our exponentials: but functions with singularities are also welcome, provided the radius of convergence of their Taylor expansions depends on the point of expansion $x_0$, and tends to infinity for $x_0 \to \infty$. A very natural natural assumption, although not the most general one, for the scattering problem we are dealing with, is that *singularities of all functions have bounded real part*. As in the previous approach, we also need, of course, that all functions decay to zero (faster than $x^{-1}$) for $x \to \infty$.

By substituting the series expansions in the equations of motion, we get, for any $h > 0$,

$$x_h(t) = \int_{-\infty}^{t} K_x(t,t') F_h(t')\,dt', \quad p_h(t) = \int_{-\infty}^{t} K_p(t,t') F_h(t')\,dt', \quad I_h(t) = \int_{-\infty}^{t} \tilde{F}(t')\,dt' ,$$

where $K_x$, $K_p$ are antisymmetric kernels,

$$K_x(t,t') = -w_{xx}(t)w_{xp}(t') + w_{xp}(t)w_{xx}(t'), \quad K_p(t,t') = -w_{px}(t)w_{pp}(t') + w_{pp}(t)w_{px}(t'),$$

and $F_h$, $\tilde{F}_h$ are given by

$$F_h(t) = \sum_{2 \le m \le h} \Phi^m(x_0(t)) \sum_{\substack{k_1,\ldots,k_m \ge 1 \\ |k|=h}} x_{k_1}(t) \cdots x_{k_m}(t)$$

$$+ g(\varphi^o + \omega t) \sum_{1 \le m \le h-1} f^m(x_0(t)) \sum_{\substack{k_1,\ldots,k_m \ge 1 \\ |k|=h-1}} x_{k_1}(t) \cdots x_{k_m}(t)$$

$$\tilde{F}_h(t) = -\gamma(\varphi^o + \omega t) \sum_{1 \le m \le h-1} V^m(x_0(t)) \sum_{\substack{k_1,\ldots,k_m \ge 1 \\ |k|=h-1}} x_{k_1}(t) \cdots x_{k_m}(t) \ .$$

For $h = 1$ such expressions should be red

$$F_1(t) = g(\varphi^o + \omega t) f(x_0(t)) \ , \qquad \tilde{F}_1(t) = \gamma(\varphi^o + \omega t) V(x_0(t)) \ .$$

¿From these relations, one naturally works out tree expansions rather similar to the previous ones; the only difference is that now one has only one type of branches, namely branches "of type $x$", but two types of vertices, depending on the choice of the term with $\Phi^m$ or with $f^m$ in the expansion for $F_h$. We shall distinguish the two possibilities by associating to any vertex $v$ a label $j_v = 0$ or $j_v = 1$, respectively. At variance with the previous case, only vertices with $j_v = 1$ now contribute to the order in $\varepsilon$, more precisely one has $h = \sum_v j_v$. An example of a tree for $x$, more precisely a tree contributing to $x_9$, is reported in figure 9; open circles and dots correspond there respectively to $j_v = 0$ and $j_v = 1$. Formal rules are that (i) any end vertex $v$ has $j_v = 1$; (ii) $m_v \ge 2$ if $j_v = 0$; (iii) trees for $I_h$ have $j = 1$ at the root vertex. Trees for $x_h$, $p_h$ and $I_h$ differ only for the root.

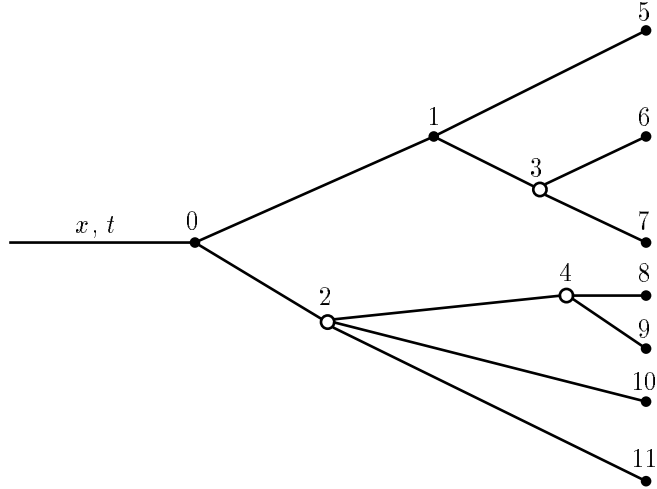

*Figure 9.* A tree with $m_0 = 2, m_1 = 2, m_2 = 3, m_3 = 2, m_4 = 2, \ldots$; open circles and dots correspond respectively to $j_v = 0$ and $j_v = 1$. The tree represents a contribution to $x_9(t)$.

As in Section 3.2, each tree with $m$ vertices represents a $m$–tuple integration over variables $\mathrm{d}t_0 \cdots \mathrm{d}t_{m-1}$, with integrals nested accordingly to the order of the tree; the function to be integrated is, here too, a product of one factor for each vertex: precisely, any vertex $v$ but the root vertex contributes with a kernel $K_x(t_{v'}, t_v)$, $v'$ denoting the vertex immediately before $v$, while the root vertex $v_0$ has kernel $K_x(t, t_{v_0})$, or $K_p(t, t_{v_0})$, or no kernel at all, respectively for $x_h$, $p_h$ or $I_h$. Moreover, any vertex with $j_v = 0$ contributes with a factor $\Phi^{m_v}(x_0(t_v))$, while, for $j_v = 1$, the factor is $g(\varphi^o + \omega t_v) f^{m_v}(x_0(t_v))$ or, for the root vertex of $I_h(t)$, $\gamma(\varphi^o + \omega t_{v_0}) V^{m_{v_0}}(x_0(t_{v_0}))$.

Everything proceeds then smoothly in the same way as with the energy–time variables, namely:

a) One makes the position: $\zeta_h = \sup_{t \in \mathbb{R}} |x_h(t)|/\beta(t)$; proceeding as in Section 3.1, it is not difficult to find, here too, a recursive relation of the form

$$\zeta_h \le \mathrm{const.} \sum_{l=1}^{h-1} \zeta_l \zeta_{h-l} \ ,$$

which assures convergence.

b) Exactly as in Section 3.2, one expands $g$ and $\gamma$ in Fourier series, and further decorates trees by indicating, at each vertex $v$ with $j_v = 1$, a Fourier index $n_v \in \mathbb{Z}$. Vertices with $j_v = 0$ have, by definition, $n_v = 0$, so the Fourier index, or momentum, of the whole tree is $\nu = \sum_v n_v$. As in Section 3.2, one raises simultaneously all integration paths, getting, for $x_h$, $p_h$ and $I_h$, and for nonvanishing momentum, exponential laws of the form (3.15).

c) Finally, to prove that the average (momentum $\nu = 0$) of $I_h$ follows the special exponential law (3.22), one produces relevant cancellations. The procedure closely follows Section 3.3, so we shall not describe it in detail. Basically, here too one introduces the decomposition (3.16), which leads to clusters, and proves exact cancellations among trees with vanishing free momentum (i.e., with Fourier component $n_0$ of the root cluster): the cancellation mechanism is, in this case, exactly the same as the one pointed out by L. Chierchia and exposed in [G1]. The equivalence classes giving exact compensations are obtained, as in the previous approach, by moving the root from one vertex with $j = 1$ to a similar one inside the root cluster; by the way, just here one uses the Hamiltonian character of the problem, namely the fact that $\gamma = -g'$ and $f = -V'$. A crucial point is that the product $K_x(t_{v'}, t_v) S(t_{v'} - t_v)$ is symmetric (in other words: the antisymmetric factor $K_x$ plays the same role as the change $\eta \leftrightarrow \xi$ on branches in the path, in the previous approach).